\documentclass[traditabstract]{aa} 
\usepackage{graphicx}
\usepackage{txfonts}
\begin{document}

\title{A unified model for the spectrophotometric development of classical and recurrent novae}
\subtitle{The role of asphericity of the ejecta}
\author{Steven N. Shore\inst{1,2,3} }

\institute{
Dipartimento di Fisica "Enrico Fermi'', Universit\`a di Pisa; \email{ivan.degennaroaquino@gmail.com; shore@df.unipi.it}
\and 
INFN- Sezione Pisa, largo B. Pontecorvo 3, I-56127 Pisa, Italy
\and
Astronomicky Ustav UK., V Holesovickach 2,180 00 Praha 8, Czech Republic
 }

\date{Received ---; accepted ---}

              \date{received ---; accepted ---}


 \abstract
 {
 
 There is increasing evidence that the geometry, and not only the filling factors, of nova ejecta is important in the interpretation of their spectral and photometric developments.  Ensembles of spectra and light curves have provided general typographies.  This Letter suggests how these can be unified.The observed spread in the maximum magnitude - rate of decline (MMRD) relation is argued to result from the range of opening angles and inclination of the ejecta, and not only to their  masses and velocities.  The spectroscopic classes can be similarly explained and linked to the behavior of the light curves.  The secondary maximum observed in some dust forming novae is a natural consequence of the asphericity.  Neither secondary ejections nor winds are needed to explain the phenomenology.   The spectrophotometric development of classical  novae can be understood within a single phenomenological model with bipolar, although not  jet-like,  mass ejecta.   High resolution spectropolarimetry will  be an essential analytical tool.}  
 

  \titlerunning{Asphericity of nova ejecta as a unifying concept} \authorrunning{S. N.
Shore}

  \maketitle

\section{Introduction}

Since the first arguments about ``island universes'',  novae have been used as standard candles for cosmography (Curtis 1917) based on their relatively small spread in absolute visual magnitude at optical maximum.   The Maximum Magnitude - Rate of Decline (MMRD) relation has been a cornerstone in nova studies since it was proposed as a distance indicator by Zwicky  and promoted by Arp (1956) and van den Bergh (1988) based on the Local Group galaxies.  The contemporary calibration traces to the study by Della Valle \& Livio (1995) who provided a parametric representation and a qualitative explanation for the relation.  In subsequent years, the MMRD has been used extensively for distance determinations for Galactic and Local Group novae.  It has also been promoted as a promising cosmological  distance calibration (Della Valle \& Gilmozzi 2002)  in the era of increasingly large telescopes and more sensitive detectors.  In what follows, the MMRD relation will be assumed as an {\it empirical} result.  The purpose of this Letter is to present a different perspective on its interpretation of its origin, which questions its unqualified utility  as a means for determining the intrinsic parameters of novae, and to link that explanation with other phenomena of the nova outburst.  In a recent paper (Shore et al. 2013) on V959 Mon = nova Mon 2012, it was shown that at least one subclass of classical novae, the ONe subgroup, is likely a standard candle.   There is also mounting evidence that nova phenomenology is best explained by non-spherical ejecta (e.g. Woudt et al. 2009, Ribeiro et al. 2011, Chesneau et al. 2012, Shore et al. 2013a,b, Ribeiro et al. 2013).  We argue that this has important consequences for the understanding of the spectrophotometric development of the outbursts.  

\section{The MMRD relation and the Tololo classification}

We begin with a basic, standard premise.  The nova event is the consequence of a thermonuclear runaway on a white dwarf (WD) that leads to the explosive ejection of both the accreted layer and, possibly, some mixed quantity of envelope material (e.g. Jos\'e 2012, Downen et al. 2013 and references therein).   Thus, the WD is not destroyed in the process and the subsequent spectroscopic and photometric phenomenology is determined by the interplay of ejecta dynamics and opacity and the post-explosion relaxation of the underlying WD.   

This is the first step in interpreting the MMRD relation.  The rise in the optical and near infrared brightness is because the ejecta act as a passive, highly wavelength dependent filter, redistributing flux from the far ultraviolet near the peak of the WD continuum into the longer wavelengths where the ejecta opacity are substantially lower.  The main opacity source, the so-called ``iron curtain''  (hereafter  called  Fe-curtain), is from relatively low ionization species of heavy elements -- mainly the Fe group because of their combined abundance and spectral complexity -- whose optically thin counterparts are in the visible and near infrared.   As shown in Shore et al. (2011), the optical P Cygni lines of these species, formed from the excited states that are pumped by the ultraviolet resonance multiplets, form at high optical depths because of a recombination front that engulfs the ejecta during the initial adiabatic cooling driven by free expansion.  Because this front is not dynamical, its speed is not limited by the hydrodynamics of the expansion and therefore the UV drop and optical rise can be very fast.  The only important factor is the rate at which the number density, $n_e$,which is a local quantity at each radial position in the ejecta, declines with time.  This, in turn, depends on the maximum velocity of the ejecta for a ballistic velocity law (one that is linear with radial distance, usually called a ``Hubble flow'').   The timescale for the recombination, varying as $(\alpha n_e)^{-1}$, where $\alpha$ is the (temperature dependent) recombination coefficient, implies that the inner regions will turn opaque rapidly, shielding the outer ejecta and accelerating the rise.   Assuming $\alpha \sim t^b$ where $b \sim 1/2$ (e.g. Osterbrock \& Ferland 2006), the recombination time scales as $t^{5/2}$ where $t$ is the time since explosion, although the exponent is not crucial to the argument.  A {\it column} density of $10^{24}$cm$^{-2}$ suffices to render the UV transitions completely opaque producing the visual maximum of the nova light curve (e.g. Hauschildt et al. 1996, Hauschildt 2008).  Thus, in this picture, all of the intercepted flux from the WD is bolometrically re-emitted at longer wavelengths.  The catch here is an implicit assumption regarding the {\it geometry} of the ejecta, that the central source is completely covered {\it not} relative to the observer but in the expanding frame.  If the ejecta are spherical, and the WD luminosity is, for instance, at some fixed value like the Eddington limit, $L_{Edd} = 3.4\times 10^4(M/M_\odot) L_\odot$, then the most energetic explosions (most luminous) should also be the brightest in the optical.  

The subsequent development of the visual photometry is determined by the rate at which the ejecta become optically thin.  This depends on the expansion velocity and the advance of a reionization front in the ejecta as the local density drops.  The shift to higher ionization states of the absorbing chemical species diminishes the efficiency of the redistribution and the optical light declines with an attendant rise in the ultraviolet (see e.g. Shore et al. 1994a, Hauschildt 2008).   The two principal parameters describing this, the time required for the optical emission to decrease by two or three magnitudes is denoted by $t_2$ and t$_3$, respectively (see the classic discussions by Payne-Gaposchkin (1957) that contains extensive references to earlier work, and Strope et al. (2010) for a new classification system).  It is interpreted in the ejecta paradigm, although not within the wind picture (e.g. Hachisu  \&  Kato 2006),  as an inevitable consequence of the local density decrease and continued irradiation of the ejecta by the WD.  Assuming sphericity,  the maximum visual magnitude should be directly related to the luminosity of the WD since the central source is completely covered.  This accounts for the ``MM'' part of the relation, while the rate of decline (the ``RD'') is governed by the expansion rate of the ejecta and decrease in the optical depth.  These should be related in a way that the maximum visual brightness is inversely proportional  to $t_2$ and/or $t_3$  in some form.  The identification of a higher expansion velocity with higher luminosity for a not too different total mass of the ejecta completes the reasoning leading to the expectation that there is an MMRD relation in the first place.

If, instead, the ejecta are {\it not} spherical, two things happen.  The first is that if the critical column density is reached, the {\it ejecta} will be opaque along any radial line of sight to the WD and, along each ray, the spectroscopic behavior will be as if the ejecta were spherical.  The photons emitted by the WD will be absorbed and convert to visual light without regard to the mass of the ejecta.  To re-iterate the point, only the local density and pathlength matter for this to occur, whether the principal mechanism is true absorption and thermalization of the photons or resonance scattering.  Thus, the radial optical depth, which is proportional to the {\it column} density, will provoke the redistribution.  But the fraction of the intercepted flux depends on the solid angle, $\Delta \Omega$,  subtended by the ejecta {\it as seen from the WD}.  If  the ejecta are aspherical, and here we will assume bipolarity in keeping with the observations (e.g. the interferometric results and the resolved ejecta such as V445 Pup, Woudt et al. 2009), only a fraction $\Delta \Omega/4\pi$ of the incident flux at radius $R(t)$ will be redistributed.  The rest will be scattered or not absorbed.  Hence, the maximum visual luminosity of the ejecta will be less than that of the central object.  Although the flux redistribution is still bolometric, it will not be conservative, the filter will not be as effective in capturing all incident photons.  We therefore describe the radiative transfer in terms of two optical depths, $\tau_\parallel$ and $\tau_\perp$ that are related by simple scaling of the opening angle of the bipolar lobes, $\Delta \theta$.  These are related as a constant ratio at any time but $\tau_\perp$ is always smaller, provided the geometry of the ejecta remains unchanged.  

The second is more subtle and relates to the taxonomic classes introduced by Williams (1992) for nova spectroscopy.  The Tololo classification is based on the overall appearance of the optical spectrum, from roughly 3500 - 8500\AA, separating the main types into ``Fe novae'' and ``He/N novae'' (Williams et al. 1991, Williams 1992).  The former show spectra at maximum and up to $t_3$ that are dominated by absorption and emission lines of the low ionization Fe group species, often displaying P Cyg lines, and including forbidden transitions.  The He/N group never display the strong Fe peak lines but are dominated by the He I/II and N 4640\AA\ complex (which includes C III and N III).  There is also a ``hybrid'' class, which transitions between these two types.  

The appearance of the Fe and related heavy metal lines is a consequence of the previous argument about the critical column density.  It depends only on the radial distance of any location in the ejecta from the WD and is {\it not} dependent on the observer.  If there is a sufficiently large radial optical depth the photons emitted from the WD will convert.  The observer, however, will see absorption only when the transverse optical depth is large enough to impede the free escape of photons.  This depends on the geometry of the ejecta.  For a spherical ejecta this is always of the same order of magnitude as the radial value.  It should be remembered that $\tau_\parallel = \tau_{radial}$ depends inversely on the velocity gradient so even though the lines are opaque there is still some residual flux that will reach an external observer.  But this is very small.  For aspherical ejecta $\tau_\perp$ may be rather large even if the column density is lower than for the radial rays because the velocity gradient is reduced by the transverse line of sight inversely as $\cos i \sin \Delta \theta$  while the column density scales linearly with the same factor where $i$ is the inclination of the symmetry axis to the line of sight.  Hence we would expect that the absorption would disappear more rapidly for strongly bipolar ejecta than for nearly spherical ones but that the emission can persist.   The same is true for the maximum magnitude, so the first prediction of this picture is the existence of novae with rapid transitions out of the absorption line Fe nova stage with relatively low maximum visual brightness.  This is moderated by the velocity of the ejecta so there should be both types of novae, with short and long decline times.  These two effects seem to be in agreement with the recent results by Cao et al. (2012) for the M31 novae observed with the Palomar Transient Factory (PTF) and GALEX.  In particular, as they note, the discovery of UV emission peaking before the optical detection would point to the asphericity of the ejecta along with the scatter obtained from their (admittedly limited) sample.\footnote{The Cao et al. results imply possible systematic observational bias in the previous samples, and highlight the large number of outliers from the original MMRD relation.  This is best seen for novae in external galaxies  because of the often large uncertainties associated with nova distances in the Galaxy. }

\section{Implications for mass estimates}

A recalcitrant problem in nova studies has been the lack of convergence of model predictions and observational determinations of ejecta masses (e.g. Jos\'e \& Shore 2008; Starrfield et al, 2012).  The disagreement has be as large as an order of magnitude but may be mitigated by adopting nonspherical geometries for the analyses.  Note that the mass is not a measured quantity.  It is derived from temperature, density, and dynamical information.  For emission lines, for instance, the emission measure $n_e^2V$, where $V$ is the volume, is the observationally accessible quantity.  To convert this into a mass requires some value for V. If the ejecta are fragmented, then $f$ is needed and can be guessed at through the fine structure on the profiles.  But if aspherical, it is not enough to know the velocity distribution in the ejecta.  Absolute expansion distances must be known (this will now depend on the inclination and shape for the maximum radial velocity) /  Furthermore,  the solid angle is not $4\pi$ and may be, depending on the specific transitions used, be considerably smaller.  Factors of four to 10 can be accounted for by such considerations.  Since the geometry can be modeled from the line profiles, there are internal consistency checks to any such mass estimate (see Shore et al. 2013a,b for applications to V959 Mon and T Pyx).

\section{Dips, cusps, and asphericity in dust forming events}

Dust formation has been known to occur in classical novae since the discovery of the deep minimum of DQ Her 1934 (see, e.g. the review by Evans \& Gehrz 2012).  It has two unmistakable signatures: a deep decline of visible flux with subsequent slow recovery to the continuation of the pre-condensation light curve, and the simultaneous rise in the near infrared flux.  The former is explained as an optically thick shell that forms when the energy density corresponds to that at the Debye temperature and then thins out over time.  The latter is due to the reprocessing of the absorbed visible and ultraviolet by the grains that are transparent in the infrared.  The emissivity, $\epsilon_\lambda$, depends on the dust composition through the absorption coefficient, $\epsilon_\lambda = \kappa_\lambda B_\lambda(T_{IR})$ where $B_\lambda(T)$ is the Planck function.  Since the radiation temperature is assumed to be due to the absorption, the same $\kappa_\lambda$ determines $T_{IR}$.   Although the theoretical picture for condensation is far from complete, we will take a more generic approach.  Based on the ultraviolet (1200-3000\AA) observations of the dust formation episode in V705 Cas 1993 (Shore et al. 1994b, Evans et al. 2005), it appears that the grains begin growing at around the same time as the UV becomes optically thin, around $t_3$, in those novae that form dust.  In the V705 Cas event, the drop in the flux shortward of 
3000\AA\ was dramatic but the spectrum remained invariant, hence the argument that the outermost parts of the ejecta spawned the grains.  For spherical ejecta, the total dust mass can be obtained from the integrated emissivity, assuming a mean grain density (composition and mean radius) from the infrared luminosity.

Dust does not simply absorb and radiate, it also screens the outer ejecta from the WD and should produce a more localized recombination wave in the gas.  The timescale for this depends on the continuum but, since this occurs at about the time the UV is increasing, the contemporary state of ionization is not large in that part.  Thus some of the emission should also come from the screened region that, however, will have a far lower density than the inner ejecta and should contribute less to the line profiles in emission,.  In absorption, however, the region will remain in this state for some time even as the dust thin and a new set of Fe peak absorption lines can appear in the optical if the escape probability based on $\tau_\parallel$ is still sufficiently large.   It should be recalled that the dust is always more effective as a screen to the UV than the optical so the lower energy photons can scatter from within the layer once the optical depth in the visible falls below about unity.   
The most recent nova to show the photometric cusp and indications of dust formation , V2362 Cyg, showed another spectroscopic feature that may support this  scenario (Munari et al. 2008, Strope et al. 2010).  The onset of the cusp was around the same time as usually seen in CO novae that are dust formers, accompanied by a strongly blue continuum and higher ionization and higher excitation emission lines.  During the event, broad absorption again appeared above the emission, which could be an example of this interplay between the cool gas and the dust.  This would be visible even after the formation of  a completely opaque  layer if the observer's line of sight is not blocked by the dust.  In spherical ejecta, the disappearance of the pseudo-photosphere means that any lines formed interior to the dust layer will be invisible.  If the cone is tilted, then both the interior (where absorption lines can form as they do in the recombination wave event) and the lines formed above the dust, will be visible.

If the ejecta are aspherical, the dust may not cover the line of sight toward the WD for an external observer but still intercept photons from both the star and from the inner ejecta.  The thermal formation condition and heating of the dust  is the same, spherical or not, and its emitted spectrum will be the same as a spherical expansion, but the luminosity  will be lower since less mass is heated {\it to the same temperature}.  But more important is that the dust forms at a specific radial distance, hence at a higher radial velocity (relative to the WD) than the emission line-forming region.  The emission lines should, therefore, reflect from a mirror that is moving outward at a significantly higher velocity.  Hence, the emission lines should suddenly broaden and strengthen since now scattered light will add to that already observed along the line of sight.  But an important testable consequence of this idea is that the emission should now be polarized from the dust scattering for the systems that show the cusp while they should be unpolarized for those showing the dip since these will be either spherical or more aligned along the line of sight.   The same is true for the continuum.  The scattered light from the pseudo-photosphere and the central WD will  increase the visual brightness proportional to the solid angle subtended by the ejecta but the scattered continuum will be polarized with a wavelength dependence that depends on the grain optical scattering phase function.   The continuum will be that of the WD and inner ejecta, hence blue, and made even bluer by the scattering (depending on the specific dust properties).  A related interpretation of infrared observations as a light echo from the old ejecta of T Pyx has been proposed by Evans et al. (2012).   

An important consequence of this picture is that there is no need to invoke a second ejection event, nor is there a need for a continuous outflow like a super-Eddington wind that has been previously used as the interpretative scheme for these events.  They should be rare since they depend on rather special observer-related conditions conditions (e.g. geometry and inclination relative to the line of sight)  but should essentially mimic the absorption event in ``reverse''.

\section{Final comments}

Other non-spherical outflows and ejecta are known among, for example, the massive stars.  Many are also dust formers.  The Luminous Blue Variables show dust formation and bipolar winds and ejecta, the most notorious example being $\eta$ Car (see e.g. Smith et al. (2011) and references therein).  Supernovae type II are also known to form dust and there is theoretical and observational evidence of asphericity in the ejection events.  Classical novae may be useful analogs to the radiative and thermal processes in these far denser and more dynamical and complex events.   In this context of massive stellar outflows, the discussion by Friedjung (2011 ) regarding the role of departures from spherical symmetry in the (hypothesized) winds during nova outbursts is particularly interesting although proposed in a contrasting paradigm to that in this paper.   Spectropolarimetry, which in nova studies has seen comparatively little application to date because of the previously low resolutions available, now holds the key.  \\
\medskip

I warmly thank Elena Mason, Ivan De Gennaro Aquino, Greg Schwarz, Daniela Korcakova, Sumner Starrfield, Jason Aufdenberg, Peter Hauschildt, Jordi Jos\'e, Massimo Della Valle, Tim O'Brien, Valerio Ribeiro, Fred Walter, Alan Shafter, and Brian Warner for many invaluable discussions and exchanges.  I also  thank Patrick Woudt for an invitation to the University of Cape Town and the conference {\it Stella Novae 2013} and for discussions during the visit, and Petr Harmanec and Marek Wolf for a visiting appointment at Charles University, where many of the ideas in this note originated.  Thanks go also to the anonymous referee whose encouragement and suggestions were most welcome.



\end{document}